\documentclass[prb,floatfix,showpacs,superscriptaddress, twocolumn]{revtex4-2}
\usepackage{graphicx}
\usepackage{amsmath}
\usepackage{braket}
\usepackage[colorlinks, linkcolor=blue, citecolor=blue, urlcolor=blue]{hyperref}
\usepackage{threeparttable}
\usepackage{textcomp}
\usepackage{siunitx}
\usepackage{dcolumn}
\usepackage{booktabs} 
\usepackage{graphicx} 
\usepackage{xcolor}
\usepackage[normalem]{ulem}
\usepackage{xurl}
\usepackage[version=4]{mhchem} 
\begin{document}

\title{Elementary magnons and interacting multi-magnon quasiparticles in the effective spin-$\frac{1}{2}$ kagome-staircase magnet \ce{Co3V2O8}}

\author{Yuan Xiao}
\affiliation{Department of Physics, University of Science and Technology of China, Hefei, Anhui 230026, People's Republic of China}

\author{Jiajun Mo}
\affiliation{Department of Physics, University of Science and Technology of China, Hefei, Anhui 230026, People's Republic of China}

\author{Zhenmeng Jiang}
\affiliation{Department of Physics, University of Science and Technology of China, Hefei, Anhui 230026, People's Republic of China}

\author{Haoyang Leng}
\affiliation{Department of Physics, University of Science and Technology of China, Hefei, Anhui 230026, People's Republic of China}

\author{Otkur Omar}
\affiliation{Department of Physics, University of Science and Technology of China, Hefei, Anhui 230026, People's Republic of China}

\author{Yanjun Li}
\affiliation{Department of Physics, University of Science and Technology of China, Hefei, Anhui 230026, People's Republic of China}

\author{Fengyi Song}
\affiliation{Department of Physics, University of Science and Technology of China, Hefei, Anhui 230026, People's Republic of China}
\affiliation{Department of Materials, School of Natural Science, The University of Manchester, Manchester M13 9PL, United Kingdom}

\author{Jianjun Ying}%
\email{yingjj@ustc.edu.cn}
\affiliation{Department of Physics, University of Science and Technology of China, Hefei, Anhui 230026, People's Republic of China}

\author{Shang Gao}%
\email{sgao@ustc.edu.cn}
\affiliation{Department of Physics, University of Science and Technology of China, Hefei, Anhui 230026, People's Republic of China}

\date{\today}

\begin{abstract}
The excitation spectrum of an anisotropic magnet provides a direct link between its microscopic Hamiltonian and interaction-driven quasiparticles. Here we use high-resolution time-domain terahertz spectroscopy to map the magnetic excitations of the three-dimensional kagome-staircase compound \ce{Co3V2O8} as functions of temperature and magnetic field. At low energies, polarization-resolved spectra identify magnetic-dipole-active one-magnon modes and track their evolution across the ferromagnetic and spin-density-wave phases. Combining their field dependence with previously reported inelastic-neutron-scattering dispersions, we determine an effective spin-$\frac{1}{2}$ Hamiltonian with strongly anisotropic exchange that quantitatively reproduces the one-magnon spectrum. This model provides a noninteracting benchmark for the high-energy response, where we observe sharp branches with field slopes that are two to four times those of the one-magnon modes, together with anticrossings between branches of different magnon numbers. Their sharpness, polarization dependence, and departure from the calculated multi-magnon continua identify them as interacting multi-magnon quasiparticles that can be stabilized by strong exchange anisotropy.

\end{abstract}
    
\maketitle
    
\section{Introduction}

Magnetic excitation spectra encode both the microscopic interactions of a quantum magnet and the collective phenomena generated by those interactions. In magnetically ordered systems, elementary one-magnon modes provide direct constraints on exchange couplings, and their low-energy dynamics are often well described by linear spin-wave theory (LSWT)~\cite{Kubo_1952_The,Oguchi_1960_Theory}. At higher energies, however, magnon-magnon interactions can redistribute spectral weight, generate multiparticle continua, and stabilize composite quasiparticles formed from two or more spin flips~\cite{Bethe_1931_Zur,Wortis_1963_Bound,Dyson_1956_General}. A unified description of the one- and multi-magnon sectors is therefore essential for distinguishing interaction-induced quasiparticles from features that follow directly from the underlying one-magnon band structure.

Experimentally, the two sectors of one- and multi-magnon excitations are often accessed through complementary spectroscopic probes. Inelastic neutron scattering (INS) resolves momentum-dependent one-magnon dispersions and is a central tool for determining microscopic magnetic~\cite{Boothroyd_2020_Principles}. Its predominantly magnetic-dipole coupling, however, generally favors single-spin-flip processes and can strongly suppress the spectral weight of higher-order excitations~\cite{Fishman_2018_Spin-Wave,Sala_2021_Van,Zoghlin_2023_Refined,Dally_2020_Three-Magnon}. Time-domain terahertz spectroscopy (TDTS), by contrast, probes the zone-center ($\Gamma$) response with controlled electric- and magnetic-field polarizations. Magnetic-dipole, electric-dipole, and magnetoelectric optical channels can make TDTS sensitive to both one-magnon modes and otherwise weak multi-magnon excitations~\cite{Koch_2023_Terahertz,Fishman_2018_Spin-Wave}. The high energy resolution of TDTS further enables sharp quasiparticle modes to be distinguished from multi-magnon continua even when their energy separation is small~\cite{Legros_2021_Observation,Peedu_2022_Terahertz,Sahasrabudhe_2020_High-field}. Combining INS and TDTS thus provides a route to establish a quantitative one-magnon baseline and to identify interaction-driven departures from noninteracting spin-wave theory.

Such departures are particularly significant in strongly anisotropic magnets. In one-dimensional Ising-like systems, effective attractive interactions can bind neighboring spin flips into stable multi-magnon states, as demonstrated in compounds such as \ce{CoNb2O6}~\cite{Woodland_2023_Tuning,Morris_2014_Hierarchy} and \ce{SrCo2V2O8}~\cite{Wang_2018_Experimental}. Repulsively bound magnons have also been observed in the quasi-one-dimensional antiferromagnet \ce{BaCo2V2O8}~\cite{Wang_2024_Experimental,Halati_2023_Repulsively}. In systems with $S>1/2$, single-ion anisotropy can additionally stabilize multiple spin flips on the same magnetic site~\cite{ElMendili_2025_Longitudinal,Lou_2024_Evolution}, as exemplified by higher-order bound states in \ce{FeI2} and related van der Waals magnets~\cite{Legros_2021_Observation,Cenker_2020_Direct,Wyzula_2022_High-Angular}. 

\begin{figure*}[ht] 
    \centering 
    \includegraphics[angle=0,width=130mm]{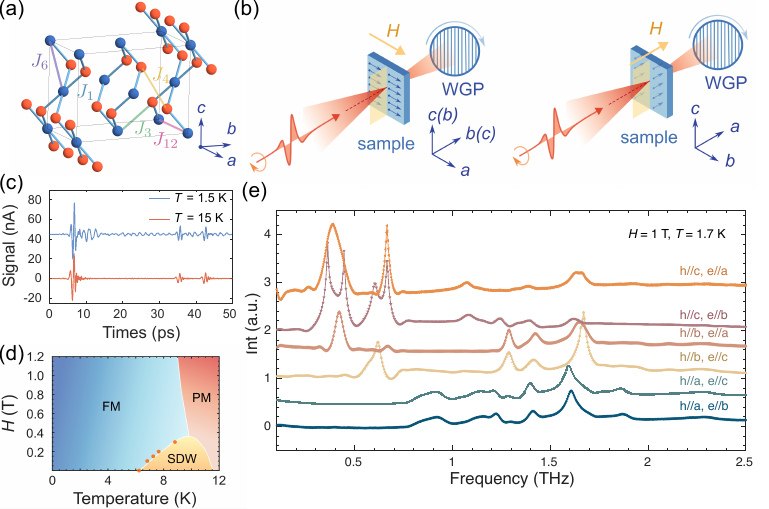}
    \caption{(a) The kagome staircase lattice formed by the \ce{Co^{2+}} ions in CVO, where the cross-tie and spine sites are shown in blue and red, respectively. The exchange paths for the couplings considered in our modeling are indicated. (b) The TDTS experimental setup in the Voigt geometry and Faraday geometry. WGP represents the wire-grid polarizer. (c) Representative zero-field time-domain terahertz signals measured at 1.5~K in the FM state (blue) and 15~K in the PM state (red). (d) Phase diagram of \ce{Co3V2O8} for $H \parallel \it{a}$, adapted from Ref.~\cite{Yen_2008_Magnetic}, where FM, PM, and SDW represent the ferromagnetic, paramagnetic, and spin density wave, respectively. The transition points between the FM and SDW phases as detected in our THz measurements are indicated by red circles. (e) Background-subtracted frequency domain absorption spectra. Different colors indicate different THz polarization configurations. The low energy single-magnon mode shows clear dependence on both the electric field $e$ and magnetic field $h$, while the high energy multi-magnon response is mainly governed by the direction of $h$.}  
    \label{fig:Introduction}
\end{figure*}

Till now, most established examples of multi-magnon quasiparticles occur in one- and two-dimensional systems~\cite{Dally_2020_Three-Magnon,Chauhan_2020_Tunable,Sheng_2025_Bose,Sahasrabudhe_2020_High-field}. In an isotropic three-dimensional system, by contrast, the exchange-mediated attraction is generally insufficient to overcome the kinetic-energy cost near the Brillouin-zone center, leading to a threshold wavevector for bound-state formation~\cite{Wortis_1963_Bound,Hanus_1963_Bound}. Strong anisotropy may relax this constraint by enhancing local spin-flip correlations, making anisotropic three-dimensional magnets a stringent setting to search for composite magnon quasiparticles~\cite{Pan_2014_Low-energy,Peedu_2022_Terahertz}.

The kagome-staircase compound \ce{Co3V2O8} (CVO) provides such a setting. Its magnetic \ce{Co^{2+}} ions occupy two inequivalent spine ($\mathsf{s}$) and cross-tie ($\mathsf{c}$) sites within a three-dimensional network, as illustrated in Fig.~\ref{fig:Introduction}(a). At low temperatures, CVO has a ferromagnetic ground state with the ordered moments oriented along the crystallographic $a$ axis. Upon warming through $T_{\mathrm{c}}\simeq 6.2~\mathrm{K}$, it enters a spin-density-wave regime characterized by a propagation vector $\mathbf{k}=(0,\delta,0)$, with $\delta$ evolving from $1/3$ to approximately $0.55$~r.l.u. as the temperature increases, before becoming paramagnetic above approximately $11.2~\mathrm{K}$~\cite{Wilson_2007_Magnetic,Chen_2006_Complex}. Previous neutron spectroscopic studies have mapped the one-magnon spectrum~\cite{Fritsch_2012_Magnetic,Helton_2020_Three-dimensional,Fritsch_2015_Quantum,Ramazanoglu_2009_Spin}, but its potential multi-magnon excitations have not been explored. Moreover, the spin-orbit-entangled \ce{Co^{2+}} ions connected through edge-sharing \ce{CoO6} octahedra naturally permit strongly anisotropic exchange interactions~\cite{Liu_2018_Pseudospin,Winter_2022_Magnetic}, which can affect both the elementary magnons and their mutual interactions.

In this work, we use high-resolution TDTS to investigate the magnetic excitation spectrum of CVO as functions of temperature and magnetic field. We first identify the low-energy one-magnon modes through their magnetic-dipole selection rules and follow their evolution across the magnetic phases and under applied field. By jointly analyzing the TDTS results and previously reported INS spectra~\cite{Helton_2020_Three-dimensional}, we establish an effective spin-$1/2$ anisotropic-exchange Hamiltonian that reproduces both the zone-center field dependence and the momentum-dependent one-magnon dispersion. This microscopic model then provides a noninteracting reference for interpreting the higher-energy response. Above approximately $1~\mathrm{THz}$, we observe sharp excitation branches with field-dependent slopes that are two to four times those of the elementary modes, in marked contrast to the broad and weak multi-magnon continua predicted without magnon-magnon interactions. Their polarization dependence and the observed anticrossings further reveal magnetoelectric activity and coupling between sectors with different magnon numbers. Our results suggest that strong exchange anisotropy can support well-defined multi-magnon quasiparticles in a three-dimensional magnet.

\section{Methods}

Polycrystalline samples of CVO were first synthesized by a conventional solid-state reaction method. Stoichiometric amounts of \ce{CoO} and \ce{V2O5} powders (3:1 molar ratio) were thoroughly mixed and sintered at 800~$^\circ$C for one day. The resulting powder was then pressed into rods with a diameter of 8~mm, which were subsequently sintered at 1050~$^\circ$C for two days to serve as feed rods. A single crystal of approximately 2~cm in length was grown using a floating-zone furnace at a rate of 0.5~mm/h~\cite{Balakrishnan_2004_Single}. The grown crystal was characterized by powder X-ray diffraction (XRD) and Laue diffraction to confirm its phase purity and orientation.

TDTS experiments were performed using a Toptica TeraFlashPro system integrated with an Oxford SpectromagPT cryomagnet. Three plate-like single-crystal samples were prepared with surface normals along the crystallographic $a$, $b$, and $c$ axes, respectively, so that the terahertz propagation vector $\mathbf{k}_{\mathrm{THz}}$ was parallel to the corresponding axis in the transmission geometry. The external magnetic field, $H$, was applied along the $a$~axis in the field-dependent measurements. Therefore, measurements on the $a$-plane sample were performed in the Faraday geometry ($\mathbf{k}_{\mathrm{THz}}\parallel H\parallel a$), giving the two polarization configurations $h\parallel b, e\parallel c$ and $h\parallel c, e\parallel b$, where $h$ and $e$ represent the magnetic and electric polarizations of the terahertz light, respectively. Measurements on the $b$- and $c$-plane samples were performed in the Voigt geometry with $\mathbf{k}_{\mathrm{THz}}\perp H$. The $b$-plane sample was used for $h\parallel c, e\parallel a$ and $h\parallel a, e\parallel c$, while the $c$-plane sample was used for $h\parallel b, e\parallel a$ and $h\parallel a, e\parallel b$. The polarization of the terahertz radiation was controlled by rotating the photoconductive antenna and wire-grid polarizers, as depicted in Fig.~\ref{fig:Introduction}(b).

The time-domain signals, exemplified by the data in Fig.~\ref{fig:Introduction}(c), were recorded within a 50~ps time window. Unless otherwise specified, the frequency-domain spectra were obtained by applying a Fast Fourier Transform (FFT) to the time-domain traces over a 25~ps tukeywin window, chosen to exclude delayed echoes arising from reflections at the sample surfaces and the cryomagnet's quartz windows. To isolate the magnetic response of the sample, the transmission spectrum at $15~$K, where CVO is in the paramagnetic state, was used as a reference for data normalization. The magnetic absorption coefficient, $\alpha(\omega, B, T)$, was calculated as:
\begin{equation}
    \alpha(\omega,B,T) = -\frac{1}{d}\ln\frac{I(\omega, B, T)}{I(\omega, B_r, T_r)} ,
\end{equation}
where $d$ is the sample thickness, $I(\omega, B, T)$ is the transmitted intensity in the frequency domain, and $B_r$ and $T_r$ are the reference field and temperature~\cite{Fishman_2018_Spin-Wave,Nagel_2013_Terahertz}.

Raman experiments were performed on the MonoVista CRS+ Raman microscope in a backscattering configuration, which is integrated with the Quantum Design OptiCool cryostat and a superconducting magnet. A 532~nm solid-state laser was used for the measurement. The laser power was maintained at 500~$\mathrm{\mu}$W to avoid heating effects. 1800~grooves/mm Bragg gratings were set in front of the spectrometer to filter the elastic scattering light.

To determine the coupling parameters in the effective spin-$\frac{1}{2}$ model for CVO, we employed a multi-objective marine predator algorithm (MOMPA)~\cite{Faramarzi_2020_Marine,Zhong_2021_MOMPA,Mo_2026_Emergent} to fit simultaneously the field-evolution of the one-magnon excitations revealed in our TDTS experiments and the dispersion of the one-magnon excitations reported in the previous INS experiments~\cite{Fritsch_2012_Magnetic,Helton_2020_Three-dimensional,Fritsch_2015_Quantum,Ramazanoglu_2009_Spin}. In evaluating the INS objective function, the uncertainty assigned to each intensity point was estimated as $c\sqrt{I_{\mathrm{INS}}}$, where $I_{\mathrm{INS}}$ is the measured INS intensity and $c$ is an overall scaling coefficient.
The magnetic excitation spectra were simulated using the \texttt{Su(n)ny}~\cite{Dahlbom_2025_Sunny} program. Using the fitted model, the cross section for magnetic absorption coefficients of one-magnon and non-interacting multi-magnon continuum excitations up to three magnons was calculated using the \texttt{spintoolkit}~\cite{SpinToolkitDevelopers_2026_SpinToolkit,Xu_2026_In} program.

\section{Overview of the terahertz spectra}

\begin{figure*}[t] 
    \centering 
    \includegraphics[angle=0,width=160mm]{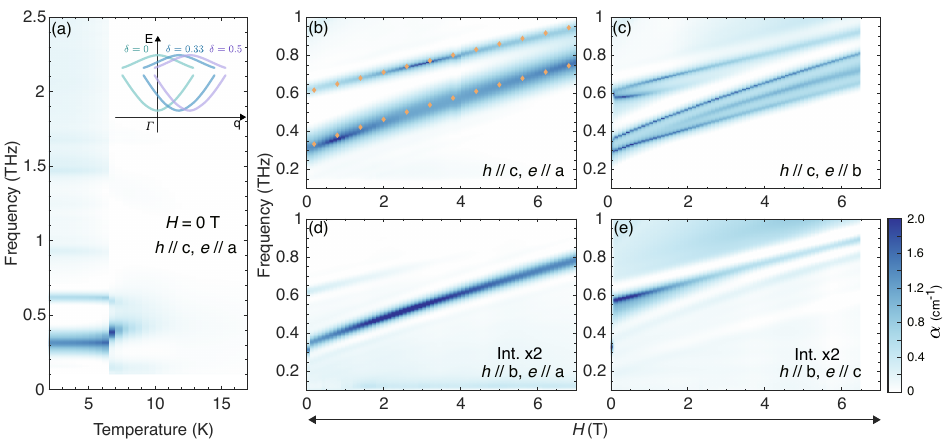}
    \caption{(a) Temperature dependence of the terahertz spectra measured with the polarization configuration of $h\parallel a$, $e \parallel c$ in zero field. A clear transition form the ferromagnetic phase to the SDW phase is observed near 6~K, followed by the gradual diminishing of the magnetic signal at higher temperatures. Inset schematically depicts the contrasting evolution of the high and low energy excitations at the ~$\Gamma$ point as $\delta$ increases from $\frac{1}{3}$ to $0.55$. (b-e) Field dependence of the low energy single-magnon mode for four THz polarization configurations: $h\parallel c, e \parallel a$ in (b), $h\parallel c, e \parallel b$ in (c), $h\parallel b, e \parallel a$ in (d), $h\parallel b, e \parallel c$ in (e). For the data in (c) and (e), the demagnetization factor is applied.  In panel (b), the orange markers indicate the magnetic field dependence of magnon modes calculated for the fitted effective spin-$\frac{1}{2}$ model.} 
    \label{fig:THz_single}
\end{figure*}

Figure~\ref{fig:Introduction}(e) presents the low-temperature magnetic excitation spectrum under different polarization configurations, with an applied magnetic field of 1~T along the $a$ axis that fully polarizes the system into the ferromagnetic state~\cite{Yen_2008_Magnetic}. Low-energy excitations in the range of [0.1, 0.8]~THz are observed for the polarization configurations of $h\parallel c$ and $h\parallel b$ regardless of the direction of $e$, but are absent for the configurations of $h\parallel a$.
This contrast follows the magnetic-dipole selection rule for the one-magnon excitations in ferromagnets, where only the transverse THz magnetic field with $h\perp a$ can drive the uniform magnon precession~\cite{Fishman_2018_Spin-Wave}. Such one-magnon interpretation is also consistent with the previously reported INS spectra~\cite{Fritsch_2012_Magnetic,Helton_2020_Three-dimensional,Fritsch_2015_Quantum,Ramazanoglu_2009_Spin}, where magnon excitations at 1.3 and 2.5 meV have been observed in zero field at $\Gamma$ and the equivalent positions~\cite{Fritsch_2012_Magnetic, Fritsch_2015_Quantum, Helton_2020_Three-dimensional}.  For fixed $h$ polarization, the exact weight of the one-magnon modes depends strongly on the THz electric-field polarization $e$, indicating additional magnetoelectric contributions to the optical matrix elements. In the range of [0.8, 2.5]~THz, rich high-energy excitations are observed, where the spectra weight exhibits remarkable similarity when $h$ is applied along the same direction. As will be discussed in section V, these high-energy modes correspond to the multi-magnon excitations that are largely magnetic dipole active.

\section{One-magnon excitations}
\subsection{Temperature evolution}

Figure~\ref{fig:THz_single}(a) exhibits the temperature dependence of the terahertz absorption spectra measured in zero field, with a polarization configuration of $h \parallel c, e\parallel a$. In this configuration, the spectrum at $T = 2.0$~K reveals a broad excitation close to $0.3~$THz and a sharp peak at $0.62~$THz, with weaker continuum-like excitations at higher energies up to $\sim2.5$~THz. A clear signature of the phase transition into the spin density wave (SDW) state is observed as an abrupt change in the magnon excitation energies. According to the previous neutron diffraction experiments~\cite{Chen_2006_Complex}, the ordering wavevector relocates from $\mathbf{q} = (0,~0,~0)$ in the FM ground state to $\mathbf{q}=(0,~1/3,~0)$ across the FM-SDW transition at $T_\mathrm{c}=6.2$~K, and then evolves towards $(0,~0.55,~0)$ with increasing temperatures. The inset of Fig.~\ref{fig:THz_single}(a) schematically illustrates the $\delta$-dependence of the $\Gamma$-point excitation energies assuming relatively weak distortion of the magnon dispersion. As the wavevector shifts from $(0,~0,~0)$ to $(0,~1/3,~0)$, a sudden renormalization of the magnon excitation energy at the $\Gamma$-point is expected, with the modes at $0.3$ and $0.62$~THz moving towards each other, both agreeing with the experiments. As $\delta$ increases from $\frac{1}{3}$ to 0.55~r.l.u., the high-energy branch continuously softens while the low-energy branch hardens.

\subsection{Magnetic field evolution}

Figures.~\ref{fig:THz_single}(b-e) present the systematic field evolution of the one-magnon excitations in CVO for fields up to 7~T applied along the $a$ axis. Data for $h\parallel a$ are omitted, since no one-magnon excitations are observed in these configurations as exemplified in Fig.~\ref{fig:Introduction}(e).  For the $h\parallel c, e\parallel b$ and $h\parallel b, e\parallel c$ configurations shown in Figs.~\ref{fig:THz_single}(c) and (e), the applied field and magnetization are perpendicular to the thin sample plate as shown in Fig.~\ref{fig:Introduction}(b),  therefore demagnetization corrections have been applied to enable comparison with the other measurement configurations.

A feature common to all polarization configurations in Figs.~\ref{fig:THz_single}(b--e) is that the one-magnon modes display an approximately linear field dependence. This agrees with the expected behavior for one-magnon excitations when the field is aligned with the ferromagnetic moment, as the energy required to flip a single spin is determined by the Zeeman energy, which scales linearly with field strength. However, the slope of the linear field dependence varies among the magnon modes. For example, in the $h\parallel c, e\parallel a$ configuration shown in  Fig.~\ref{fig:THz_single}(b), the mode near $0.3$~THz has a slightly steeper slope than the one near $0.62$~THz, indicating different $g$-factors for the $\mathsf{s}$ and $\mathsf{c}$ sites.

Comparing the spectra for the $h\parallel c$ configurations in Figs.~\ref{fig:THz_single}(b,c) with the corresponding $h\parallel b$ configurations in Figs.~\ref{fig:THz_single}(d,e) reveals weaker spectral weight in the latter, despite their similar field evolution. This difference can be attributed to the anisotropic susceptibility of transverse magnon fluctuations along the $c$ and $b$ axes, as will be discussed in the modeling subsection. More pronounced differences appear when comparing the spectra of the $a$-plane sample shown in Figs.~\ref{fig:THz_single}(c) with those of the $b$- and $c$-plane samples shown in Figs.~\ref{fig:THz_single}(b,d). In the $a$-plane sample, the applied field strongly splits the magnon excitations. As demonstrated in Fig.~\ref{fig:THz_Raman} in the Appendix, the Raman spectra of phonon excitations in CVO show no sharp anomalies across the phase transitions, ruling out structural distortion as the origin of the observed magnon splitting at low temperatures.

\begin{figure}[ht] 
    \centering 
    \includegraphics[angle=0,width=85mm]{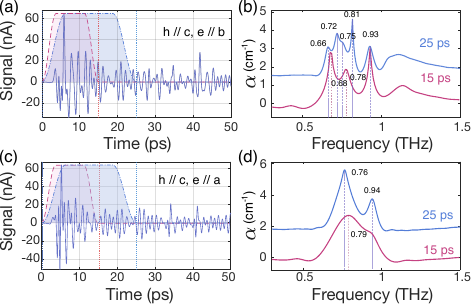}
    \caption{(a) Time-domain signals for the $h\parallel c,e\parallel b$ configuration measured in a field of 7~T. (b) The corresponding frequency-domain spectra obtained after applying time windows of 15 and 25~ps. (c,d) Similar time- and frequency-domain spectra for the $h\parallel c,e\parallel a$ configuration measured in a field of 7~T. }
    \label{fig:THz_timewin}
\end{figure}

To establish the extrinsic origin of the magnon splitting in the $a$-plane sample, Fig.~\ref{fig:THz_timewin} compares the TDTS spectra at 7~T obtained with different FFT time windows. As revealed in Figs.~\ref{fig:THz_timewin}(a,b) for the $h\parallel c, e\parallel b$ configuration, the energies of the split bands depends strongly on the FFT window size: the four peaks at 0.66, 0.72, 0.75, and 0.81~THz obtained with a 25~ps merge into two peaks at 0.68 and 0.78~THz when the window is reduced to 15~ps. In contrast, for the $h\parallel c, e\parallel a$ configuration shown in Figs.~\ref{fig:THz_timewin}(c,d), the number of peaks remains unchanged despite window-dependent broadening. The window dependence of the $a$-plane sample suggests that the pronounced peak splitting arises from a geometry-specific electrodynamic reshaping of the sharp magnetic resonance. In this geometry, with $\mathbf{k}_{\mathrm{THz}}$ parallel to the magnetization, the incident linearly polarized THz wave can be decomposed into left- and right-circularly polarized components, which may acquire different phases and losses upon interacting with the one-magnon resonance. Internal reflections or echoes within the platelet can therefore modify the local line shape near the magnetic resonance, producing the apparent multi-peak structure~\cite{Ling_1994_Theoretical,Sycz_2010_Resonant}.

\subsection{Microscopic model for the one-magnon excitations}

\begin{table}[t!]
\caption{Parameters of the Hamiltonian derived from combined fits of INS and TDTS spectra. The exchange matrices $J_1$, $J_3$, $J_4$, and $J_6$ are defined over the site pairs of $(1, 10)$, $(1, 9)$, $(1, 5)$, and $(9, 11)$, respectively, where the atomic indices correspond to the \ce{Co} ions defined in Table~\ref{tab:lattice_coords} in the Appendix. The magnetic axes ${x, y, z}$ are defined along the crystallographic directions ${a, b, c}$, respectively. The DM contribution is included in the antisymmetric part of the listed exchange matrices. All values are in meV, except for the dimensionless effective $g$-factors, $g_{\mathsf{s}}^{x}$ and $g_{\mathsf{c}}^{x}$. Values in parentheses denote one standard deviation. Uncertainties in the exchange parameters were estimated from the standard deviations of 200 parameter sets across the Pareto front, allowing for $0.30\%$ variations in $\chi^2_{\mathrm{INS}}$. Those in the $g$-factors are upper limits derived by propagating the TDTS peak-position errors through a weighted fit.}
\label{tab:parameters}

\renewcommand{\arraystretch}{1.5}

\begin{ruledtabular}
\begin{tabular}{
    c
    S[table-format=1.2(1)]
}
Parameter & \multicolumn{1}{c}{This work} \\
\colrule

$J_1$ &
\multicolumn{1}{c}{
$\left(
\begin{array}{*{3}{S[table-format=-1.2(1)]}}
 0.56(4) &  0       &  0        \\
 0       &  0       & -0.39(3)  \\
 0       & -0.24(1) & -1.57(1)
\end{array}
\right)$
} \\[2ex]

$J_3$ &
\multicolumn{1}{c}{
$\left(
\begin{array}{*{3}{S[table-format=-1.2(1)]}}
-2.02(4) & 0       &  0 \\
 0       & 0       &  0.19(1) \\
 0       & 0.24(1) & -0.60(1)
\end{array}
\right)$
} \\[2ex]

$J_4$ &
\multicolumn{1}{c}{
$\left(
\begin{array}{*{3}{S[table-format=-1.2(1)]}}
-3.00(1) &  0 & 0 \\
 0       &  0.71(1) & -1.64(1) \\
 0       & -1.64(1) &  0
\end{array}
\right)$
} \\[2ex]

$J_6$ &
\multicolumn{1}{c}{
$\left(
\begin{array}{*{3}{S[table-format=-1.2(1)]}}
 0.10(1) &  0 &  0 \\
 0       &  0.10(1) &  0.08(1) \\
 0       & -0.08(1) &  0.10(1)
\end{array}
\right)$
} \\[2ex]

$J_{12}$              & 0.09(1) \\
$g_{\mathsf{s}}^{x}$  & 4.80(7) \\
$g_{\mathsf{c}}^{x}$  & 3.41(1) \\

\end{tabular}
\end{ruledtabular}
\end{table}

\begin{figure}[bt!] 
    \centering 
    \includegraphics[angle=0,width=85mm]{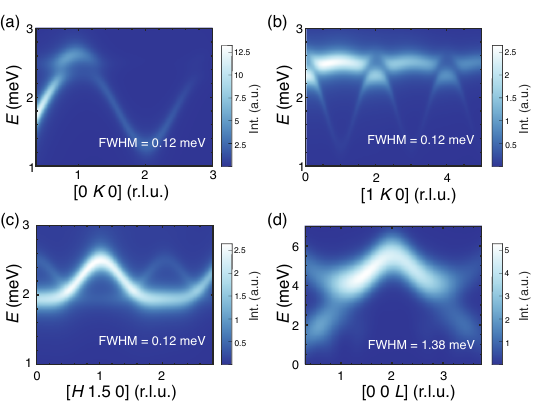}
    \caption{(a-d) Calculated inelastic neutron-scattering spectra along the (a) [0 $K$ 0], (b) [1 $K$ 0], (c) [$H$ 1.5 0], and (d) [0 0 $L$] reciprocal-space cuts based on the fitted effective spin-$1/2$ exchange model. The calculated spectra were convolved by a Gaussian function with full-width-at-half-maximum (FHWM) indicated in each panel.}
    \label{fig:INS}
\end{figure}

\begin{figure*}[b!ht!] 
    \centering \includegraphics[angle=0,width=150mm]{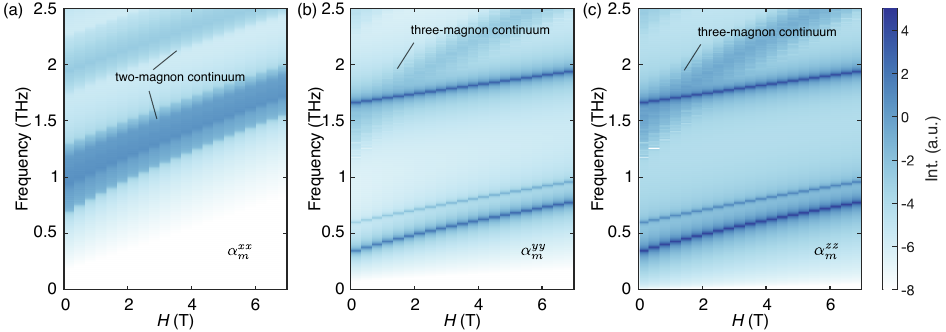}
    \caption{(a-c) Calculated field-dependent magnetic-dipole absorption coefficients (a) $\alpha_{\mathrm{m}}^{xx}$, (b) $\alpha_{\mathrm{m}}^{yy}$, and (c) $\alpha_{\mathrm{m}}^{zz}$ for THz magnetic fields oriented along the $a$, $b$, and $c$ axes, respectively. For better visibility of the weak magnon excitations, the intensity is plotted on a logarithmic scale. Multi-magnon continua are indicated in each panel.}
    \label{fig:INS-field}
\end{figure*}
For quantitative analysis of the intrinsic one-magnon excitations in CVO, we resort to the linear spin wave theory. The magnetic \ce{Co^{2+}} ions in CVO reside in a distorted octahedral environment~\cite{Rai_2007_High-energy}. 
The interplay between the crystal field and spin-orbit coupling splits the \ce{Co^{2+}} states into $J_\mathrm{eff}=1/2$, $3/2$, and $5/2$ manifolds~\cite{Lines_1963_Magnetic}. Through point-charge calculations implemented in the \texttt{PyCrystalField} package~\cite{Scheie_2021_PyCrystalField} assuming a spin-orbit coupling strength of $\lambda$ to $-18.7$~meV~\cite{Kant_2008_Optical}, we find that the ground-state Kramers doublet is well separated from the higher quartet by a gap of $\sim$30~meV, which suggests an effective spin-$\frac{1}{2}$ description of the low-energy magnetic dynamics. At the $\mathsf{c}$ and $\mathsf{s}$ sites, the calculated $g$-tensors in the crystallographic $(a,b,c)$ coordinate system are $g_c=(3.04, 2.24, 8.27)$ and $g_s=(6.76, 3.96, 2.85)$, respectively, reflecting the strong site-dependent spin-orbit-entangled crystal-field environment.

Although previous INS spectra of CVO were described using a dipole model with single-ion anisotropy~\cite{Helton_2020_Three-dimensional,Ramazanoglu_2009_Spin}, such a description is not unique. In the effective spin-$\frac{1}{2}$ subspace, the effects of the low-symmetry crystal field are more naturally encoded in anisotropic exchange interactions rather than in single-ion anisotropy~\cite{Liu_2018_Pseudospin,Winter_2022_Magnetic}. We therefore construct an effective spin-$\frac{1}{2}$ Hamiltonian by extending the isotropic exchanges in the dipole model~\cite{Helton_2020_Three-dimensional,Ramazanoglu_2009_Spin} to anisotropic exchanges
\begin{equation}
    \mathcal{H} = \sum_{\substack{\langle i,j \rangle \in 1,3,4,6 \\ \mu,\nu \in x,y,z}} 
    S_i^{\mu} J_{ij}^{\mu \nu} S_j^{\nu} + \sum_{\langle i,j \rangle \in 12} J_{ij} \mathbf{S}_i \cdot \mathbf{S}_j .
    \label{eq:ham}
\end{equation}
In this Hamiltonian, the exchange couplings include anisotropic $J_1$, $J_3$, $J_4$, and $J_6$ over the first, third, fourth, and sixth neighbors, respectively, and isotropic $J_{12}$ over the twelfth neighbors, all shown in Fig.~\ref{fig:Introduction}(a). The magnetic axes ${x, y, z}$ are defined along the orthorhombic crystallographic directions ${a, b, c}$, respectively. The anisotropic interactions are defined by the exchange matrix:
    $$\mathcal{J}_n = \begin{pmatrix} J_n^{xx} & 0 & 0 \\ 0 & J_n^{yy} & J_n^{yz}+D_n^x \\ 0 & J_n^{yz}-D_n^x & J_n^{zz} \end{pmatrix},$$
where the symmetric off-diagonal exchange, $J_n^{yz}$, and the $x$ component of the Dzyaloshinskii-Moriya (DM) interaction, $D_n^x$, are contained in the $yz$ and $zy$ matrix elements. The $J_n^{xy}$ and $J_n^{xz}$ components are not sufficiently constrained by the one-magnon dispersion and are set to zero in the present fits. For $J_4$, $D_4^{x}=0$ as dictated by symmetry. For $J_6$, the symmetry imposes $J_6^{yz}=0$.

By jointly analyzing the TDTS data and the comprehensive  INS spectra reported in Ref.~\cite{Helton_2020_Three-dimensional}, we extracted the exchange interaction parameters as summarized in Table~\ref{tab:parameters}. Marginal parameters, including $J_1^{yy}$, $J_3^{yy}$, and $J_4^{zz}$, are assigned to zero, and equal diagonal exchange parameters $J_6^{xx} = J_6^{yy} = J_6^{zz}$ over $J_6$ are assumed for simplification. In our fits, the TDTS data constrain the field evolution of the one-magnon excitation energy at the Brillouin zone center, while the INS spectra constrain the magnon dispersion across the Brillouin zone in zero field. The calculated field evolution of the one-magnon energy is compared with the TDTS data shown in Fig.~\ref{fig:THz_single}(b). Representative INS spectra calculated using the fitted model are shown in Figs.~\ref{fig:INS}(a-d), which can be directly compared to the experimental report in Ref.~\cite{Helton_2020_Three-dimensional}. 

From our fits, the effective $g$-factors for fields along the $x$ direction at the $\mathsf{s}$ and $\mathsf{c}$ sites are determined to be $g_\mathsf{s}^{x} = 4.80(7)$ and $g_\mathsf{c}^{x} = 3.41(1)$, respectively. These values are consistent with the ordered moment magnitudes determined from neutron diffraction~\cite{Chen_2006_Complex} and follow the same site-dependent hierarchy suggested by the point-charge calculations. Similar to the role played by the $a$- and $c$-axis single-ion anisotropies in the dipole model~\cite{Helton_2020_Three-dimensional}, the strongly anisotropic diagonal exchange components are essential for reproducing the gap and bandwidth of the one-magnon dispersion. In particular, the large ferromagnetic $J_3^{xx}$ and $J_4^{xx}$ terms stabilize the $a$-axis ferromagnetic ground state, while their strong contrast with the $yy$ and $zz$ components reveals sizable exchange anisotropy. The off-diagonal $yz/zy$ matrix elements, containing both symmetric anisotropic exchange and the antisymmetric DM contribution, produce the band hybridization near $\sim2.5$~meV. The DM component inferred from the antisymmetric part of the $J_1$ matrix is $\sim0.07(2)$~meV, notably smaller than the value of 0.36(9) proposed in the previous dipole model after scaling for the spin magnitude~\cite{Helton_2020_Three-dimensional}.

 Using the fitted model, we calculate the magnetic-dipole contribution to the absorption coefficient probed by TDTS~\cite{Fishman_2018_Spin-Wave}:
\begin{equation}
    \alpha^{\alpha\beta}_{\mathrm{m}}(\omega) \propto \omega h^{\alpha} h^{\beta} S^{\alpha\beta}(\mathbf{q} = 0, \omega),
    \label{eq:Smm}
\end{equation}
where $S^{\alpha\beta}(\mathbf{q}, \omega)$ with $\alpha, \beta = x,y,z$ is the magnetic response function calculated from the fitted spin model. For the non-interacting multi-magnon cross section within linear spin wave theory, we utilized the \texttt{spintoolkit} program~\cite{SpinToolkitDevelopers_2026_SpinToolkit,Xu_2026_In}. Figures~\ref{fig:INS-field} present the calculated absorption coefficients $\alpha_{m}^{xx}$, $\alpha_{m}^{yy}$, and $\alpha_{m}^{zz}$, which correspond to the experimental configurations of $h \parallel a$, $b$ and $c$, respectively. Consistent with the experimental observations, no one-magnon excitations are observed for $h \parallel a$, and the calculated spectral weight for $h \parallel b$ is markedly weaker than that for $h \parallel c$ due to anisotropic susceptibility of the magnon fluctuations. The one-magnon excitation at $\sim1.6$~THz, which corresponds to the band top of the magnon dispersion shown in Fig.~\ref{fig:INS}(d), is not resolved in our TDTS experiments, possibly because its electric-dipole transition matrix elements interfere destructively and thereby cancel its spectral weight.

\section{Multi-magnon excitations}

\subsection{Emergence of multi-magnon quasiparticles}

Besides the one-magnon excitations, our non-interacting calculations predict weak multi-magnon continua, as indicated in Fig.~\ref{fig:INS-field}. For a ferromagnet of U(1) spin-rotation symmetry about the ordered $x$ axis, $S^x$ is conserved. Consequently, a single-spin magnetic operator can change the total $S^x$ quantum number by at most one unit, and its matrix element between the fully polarized ground state and a genuine multi-magnon state with $|\Delta S^x|>1$ vanishes exactly. The multi-magnon continua obtained in our calculations can therefore be attributed to the breaking of this U(1) symmetry by the spin Hamiltonian in Eq.~(\ref{eq:ham}). In particular, the inequivalence of the $J^{yy}$ and $J^{zz}$ exchange couplings, together with nonzero symmetric off-diagonal $J^{yz}$ components, breaks continuous spin-rotation symmetry while preserving magnon-number parity. As a result, even- and odd-magnon sectors remain distinct: the longitudinal magnetic response couples to even-magnon excitations, whereas the transverse response couples to odd-magnon excitations. This accounts for the appearance of two- and three-magnon continua in the longitudinal and transverse susceptibilities, respectively.

In contrast to the calculations, our TDTS experiments reveal qualitatively different high-energy responses in CVO. As already shown in Fig.~\ref{fig:Introduction}(e), in addition to the diffuse broad features expected for multi-magnon continua, sharp excitation peaks with intensities comparable to those of the single-magnon modes are observed. These sharp high-energy excitations are particularly pronounced in the $h \parallel b$ polarization configurations~\cite{Sala_2021_Van}.
Therefore, we present in Figs.~\ref{fig:multi-mag}(a,b) the field-dependent high-energy terahertz spectra for the $h \parallel b, e\parallel c$ and $h \parallel b, e\parallel a$ configurations. Notably, in the energy regime above 1~THz, several distinct excitation modes emerge. As  labeled in the figure, the field-dependent slopes of these modes are approximately 2 to 4 times steeper than those of the fundamental single-magnon excitations. In the upper-left region, the spectral intensity has been multiplied by a factor of 2 to enhance the visibility of 4-magnon modes. The spectral sharpness and high slope of these high-energy modes provide compelling evidence for the appearance of  multi-magnon quasiparticles in CVO.

To more clearly resolve the polarization dependence of the multi-magnon quasiparticles, we extracted the peak positions of the excitation branches for the $h\parallel b, e\parallel c$ and $h\parallel b, e\parallel a$ configurations, which are overplotted in Fig.~\ref{fig:multi-mag}(c). Remarkably, with the exception of the 3m-1 branch, the peak positions of all other modes coincide perfectly, revealing marginal impacts of the sample geometry on the multi-magnon excitations. This excellent agreement, together with the strong variances for different $h$ directions shown in Fig.~\ref{fig:Introduction}(e), indicates that the multi-magnon excitations are predominantly governed by magnetic dipole transitions instead of the electric dipole transitions. In stark contrast, the 3m-1 branch exhibits finite spectral weight exclusively in the $e \parallel c$ configuration. This unique polarization selectivity suggest the possible activity of the electric dipoles for selected multi-magnon modes.

Within the effective spin-$\frac{1}{2}$ description, the same exchange anisotropy that stabilizes the $a$-axis ferromagnetic ground state may also induce attractive interactions in the multi-magnon sector. The fitted exchange matrices in Table~\ref{tab:parameters} contain dominant ferromagnetic longitudinal components, $J_3^{xx}=-2.02(4)$~meV and $J_4^{xx}=-3.00(1)$~meV. An isolated spin flip creates energetically costly antiparallel bonds along the ordered-moment direction. When two spin flips occupy neighboring sites connected by a $J_3$ or $J_4$ bond, their shared bond remains ferromagnetically aligned, so the exchange-energy cost is lower than that of two spatially separated magnons~\cite{ElMendili_2025_Longitudinal}. This broken-bond mechanism produces a short-range attraction between magnons~\cite{Wortis_1963_Bound,Dong_2023_Two-magnon,ElMendili_2025_Longitudinal}. The fitted anisotropic-exchange model therefore provides a microscopic explanation of the sharp multi-magnon quasiparticles in CVO, although a quantitative determination of their binding energies and spectral weights requires calculations beyond the LSWT scheme.

\subsection{Hybridization among the multi-magnon quasiparticles}

\begin{figure*}[ht!] 
    \centering 
    \includegraphics[angle=0,width=160mm]{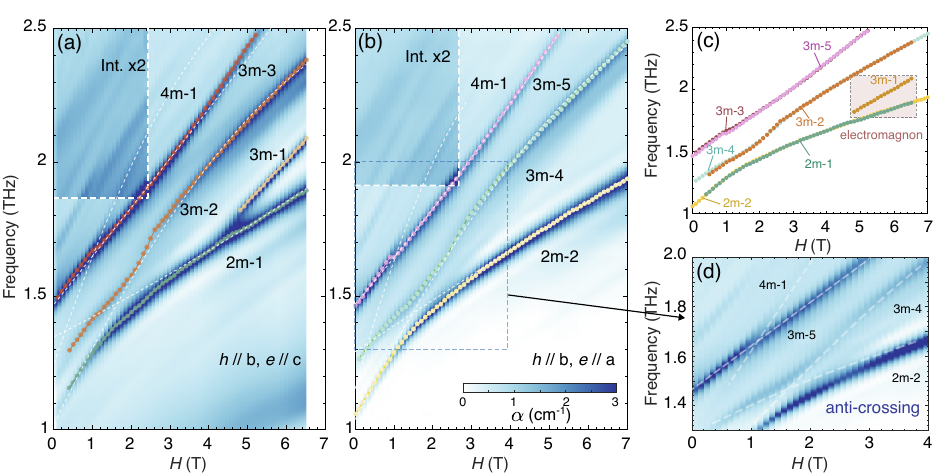}
    \caption{(a,b) Field dependence of the multi-magnon excitations measured in the (a) $h\parallel b, e\parallel c$ and (b) $h\parallel b, e\parallel a$ configurations. Demagnetization correction has been applied for $h\parallel b, e\parallel c$ in panel (a).  The main excitation branches are marked by colored circles, and the white dashed lines indicate the expected trajectories of uncoupled multi-magnon modes. Intensities in the upper left region are multiplied by a factor of 2 for visibility of the 4-magnon modes. (c) Peak positions of the main modes extracted from the $h\parallel b, e\parallel c$ and $h\parallel b, e\parallel a$ data. The 3m-1 mode between 1.7 and 2.1~THz appears only for $e\parallel c$, identifying it as an electromagnon mode. (d) Enlarged view of panel (b), emphasizing the anticrossing between the two-magnon and three-magnon modes, as well as the anomalous broadening near the crossing of the three-magnon and four-magnon modes.} 
    \label{fig:multi-mag}
\end{figure*}

Besides the observation of the multi-magnon quasiparticles, our TDTS measurements of CVO further reveal signatures of hybridization among the multi-magnon branches. In Figs.~\ref{fig:multi-mag}(a,b) and in the enlarged view in Fig.~\ref{fig:multi-mag}(d), white dashed lines represent quadratic extrapolations of the uncoupled multi-magnon modes that are obtained from fits to the Lorentzian peak centers outside the hybridization regions. Near 1.5~THz, the 2m-2 and 3m-4 branches exchange slopes, forming an avoided crossing accompanied by a modest increase in linewidth. A second anomaly occurs near 1.65~THz, where the 3m-5 and 4m-1 branches approach one another. Near this crossing, the spectra broaden, and the spectral weight is transferred from the 3m-5 to the 4m-1 branch in fields above the crossing point. These observations provide evidence for hybridization among the multi-magnon quasiparticles.

The observed hybridization between states with even and odd magnon numbers indicates that the magnon-number parity is not an exact symmetry of the system. Equivalently, the Hamiltonian must contain terms with odd changes in the magnon number. With $x$ as the spin-quantization axis, the symmetry-allowed $J^{xy}$ and $J^{xz}$ exchanges contain a single transverse ladder operator $S^\pm$. They can therefore change the magnon number by an odd integer and directly couple even- and odd-magnon states. Although the $J^{xy}$ and $J^{xz}$ exchanges have little impact on the harmonic one-magnon dispersion and are consequently omitted from our minimal model in Eq.~(\ref{eq:ham}), they may appear at higher orders in the spin-wave expansion and play an important role in the hybridization of multi-magnon excitations.

\section{Conclusion}

In summary, we have used high-resolution TDTS to investigate the magnetic excitations of the three-dimensional kagome staircase compound \ce{Co3V2O8}. The low-energy modes follow the magnetic-dipole selection rule for single magnons in the ferromagnetic state, while their spectral weights are strongly modulated by the THz electric-field polarization, revealing sizable magnetoelectric optical activity. At higher energies, we observe sharp field-dependent excitation branches whose slopes are several times larger than those of the single-magnon modes, identifying them as interacting multi-magnon quasiparticles rather than simple continuum features.

By combining the TDTS data with previously reported INS spectra, we establish an effective spin-$\frac{1}{2}$ anisotropic exchange model that reproduces the one-magnon dispersion and provides a non-interacting LSWT benchmark. The failure of this benchmark to account for the sharp high-energy branches, together with the observed inter-mode coupling, points to strong magnon interactions driven by anisotropic exchanges. Our results demonstrate that multi-magnon quasiparticles can be stabilized in a three-dimensional magnet and highlight TDTS as a sensitive probe of many-body excitation channels.


\section{Acknowledgments}
We acknowledge helps from Vilmos Kocsis, S\'andor Bord\'acs, and Naoki Ogawa for the initial setup of the terahertz spectroscopy system. We also acknowledge helpful discussions with Zhentao Wang and Haiqing Lin. This work was supported by National Key R\&D Program of China under the Grant No.~2024YFA1613100 and the National Natural Science Foundation of China (NSFC) under the Grant No.12374152.

\section{Appendix}

\begin{figure}[h] 
    \centering 
    \includegraphics[angle=0,width=80mm]{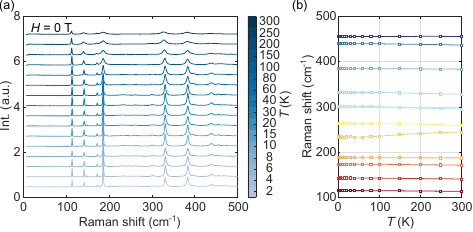}
    \caption{
  (a) Unpolarized temperature-dependent Raman spectra of CVO. The characteristic phonon modes of CVO exhibit no discernible changes with decreasing magnetic temperature (b).}  
    \label{fig:THz_Raman}
\end{figure}

Figure~\ref{fig:THz_Raman} summarizes the Raman spectra for \ce{Co3V2O8}. The characteristic phonon peaks exhibit no discernible splitting across $T_{\text{N}} = 11.2~\text{K}$, indicating the absence of structural distortion across the magnetic transition. Therefore, the splitting of the 0.32~THz peak revealed in the TDTS experiments is not due to structural distortion.

\begin{table}[h!]
\centering
\caption{Fractional coordinates of the 12 Co sites within the unit cell ($Cmce$, No. 64). Co$_c$ and Co$_s$ represent the cross-tie and spine sites, respectively.}
\label{tab:lattice_coords}
\begin{tabular}{@{}ccccccc@{}}
\toprule
Index & Ion Label &   $x$ & $y$ & $z$ \\ \midrule
1     & Co\textsubscript{s}               & 0.25 & 0.3685 & 0.25 \\
2     & Co\textsubscript{s}              & 0.75 & 0.3685 & 0.25 \\
3     & Co\textsubscript{s}              & 0.25 & 0.8685 & 0.25 \\
4     & Co\textsubscript{s}             & 0.75 & 0.8685 & 0.25 \\
5     & Co\textsubscript{s}             & 0.25 & 0.1315 & 0.75 \\
6     & Co\textsubscript{s}               & 0.75 & 0.1315 & 0.75 \\
7     & Co\textsubscript{s}               & 0.25 & 0.6315 & 0.75 \\
8     & Co\textsubscript{s}             & 0.75 & 0.6315 & 0.75 \\ \addlinespace
9     & Co\textsubscript{c}            & 0.00 & 0.0000 & 0.00 \\
10    & Co\textsubscript{c}             & 0.50 & 0.5000 & 0.00 \\
11    & Co\textsubscript{c}              & 0.50 & 0.0000 & 0.50 \\
12    & Co\textsubscript{c}            & 0.00 & 0.5000 & 0.50 \\ \bottomrule
\end{tabular}
\end{table}

The magnetic lattice used for linear spin-wave theory (LSWT) simulations was constructed based on the orthorhombic space group $Cmce$ (No.~64). The lattice parameters are defined as $a = 5.95$~\AA, $b = 11.47$~\AA, and $c = 8.28$~\AA. The unit cell comprises 12 magnetic Co ions, with fractional coordinates listed in Table~\ref{tab:lattice_coords}.

\bibliography{myref_cvo}

\end{document}